\documentclass[conference]{IEEEtran}
\IEEEoverridecommandlockouts
\usepackage{amsmath,amsfonts}
\usepackage{algorithm}
\usepackage{algpseudocode} 
\usepackage{tabularx}
\usepackage{array}
\usepackage{caption}
\usepackage{subcaption}
\usepackage{textcomp}
\usepackage{stfloats}
\usepackage{url}
\usepackage{verbatim}
\usepackage{graphicx}
\usepackage{cite}
\usepackage{csquotes}
\usepackage[table]{xcolor}
\usepackage{pgfpages}
\pgfpagesuselayout{resize to}[a4paper]

\def\BibTeX{{\rm B\kern-.05em{\sc i\kern-.025em b}\kern-.08em
    T\kern-.1667em\lower.7ex\hbox{E}\kern-.125emX}}
\begin{document}

\title{Cooperative Resource Trading for Network Slicing in Industrial IoT: A Multi-Agent DRL Approach\\
	\thanks{This work is supported in part by the Natural Science Foundation of China under Grant No.61806040 and Grant No. 61771098; in part by the fund from the Department of Science and Technology of Sichuan Province under Grant No. 2020YFQ0025; in part by the Natural Science Foundation of Guangdong Province under Grant No. 2021A1515011866; in part by the Social Foundation of Zhongshan Sci-Tech Institute under Grant 420S36; and in part by the fund from Intelligent Terminal Key Laboratory of Sichuan Province under Grant No. SCITLAB-1018.}
}

\author{\IEEEauthorblockN{1\textsuperscript{st} Gordon Owusu Boateng}
	\IEEEauthorblockA{\textit{School of Computer Science and Engineering} \\
		\textit{University of Electronic Science and Technology of China}\\
		Chengdu, China \\
		boatenggordon48@gmail.com}
	\and
	\IEEEauthorblockN{2\textsuperscript{nd} Guisong Liu}
	\IEEEauthorblockA{\textit{School of Computing and Artificial Intelligence} \\
		\textit{Southwestern University of Finance and Economics}\\
		Chengdu, China \\
		gliu@swufe.edu.cn}
}

\maketitle

\begin{abstract}
The industrial Internet of Things (IIoT) and network slicing (NS) paradigms have been envisioned as key enablers for flexible and intelligent manufacturing in the industry 4.0, where a myriad of interconnected machines, sensors, and devices of diversified quality of service (QoS) requirements coexist. 
To optimize network resource usage, stakeholders in the IIoT network are encouraged to take pragmatic steps towards resource sharing. However, resource sharing is only attractive if the entities involved are able to settle on a fair exchange of resource for remuneration in a \textit{win-win} situation. 
In this paper, we design an economic model that analyzes the multilateral strategic trading interactions between sliced tenants in IIoT networks. We formulate the resource pricing and purchasing problem of the seller and buyer tenants as a cooperative Stackelberg game. Particularly, the cooperative game enforces collaboration among the buyer tenants by coalition formation in order to strengthen their position in resource price negotiations as opposed to acting individually, while the Stackelberg game determines the optimal policy optimization of the seller tenants and buyer tenant coalitions. 
To achieve a Stackelberg equilibrium (SE), a multi-agent deep reinforcement learning (MADRL) method is developed to make flexible pricing and purchasing decisions without prior knowledge of the environment. 
Simulation results and analysis prove that the proposed method achieves convergence and is superior to other baselines, in terms of utility maximization.
\end{abstract}


\section{Introduction}
The emerging industrial Internet of Things (IIoT) paradigm is envisioned to revolutionize the industry 4.0 by supporting the interconnection of machines, devices, and servers,
to enhance productivity \cite{9732420}. However, it is challenging for key IIoT services such as smart factory, smart energy, and smart transportation, with differentiated quality of service (QoS) requirements to coexist in the same network at the same time. \textit{Network slicing (NS)} has emerged as a viable solution to address this crucial challenge in IIoT by accommodating diverse services on a common physical infrastructure, thanks to software defined networking (SDN) and network function virtualization (NFV) \cite{9629333}. In NS, the physical infrastructure (and resources) of the mobile network operator (MNO) are abstracted, partitioned, and isolated into independent virtualized networks (and resources), and assigned to \textit{sliced tenants} for their individual management. To ensure efficient resource optimization, the resources of the tenants can be adjusted according to their real-time changing service requirements.

Most existing literature on NS in IIoT primarily emphasize on its technical functionality, without a clear strategy definition or template for its economic trading benefits \cite{9498080, 9686109, 8362975}. 
For instance, the authors in \cite{9498080} designed an SDN-based architecture for dynamic slice admission and resource reservation in IIoT. Umagiliya \textit{et al.} \cite{9686109} discussed how different NS strategies can be employed in a smart factory environment, by focusing on network statistics such as bandwidth utilization and number of connected clients. 
Game theory has emerged as an analytical tool for modeling the business strategic interactions between buyers and sellers to achieve optimal and fair trading strategies \cite{9646249}. The work in \cite{9763379} formulated a non-cooperative game model to enable MEC nodes acquire virtual CPU resources from a centralized MEC orchestrator. Nonetheless, the presence of a single CPU resource provider violates competition or collaboration, which in turn creates a monopolistic market. 
With this in mind, Jiang \textit{et al.} \cite{9509582} modeled the IIoT data sharing interactions between multiple data owners and edge devices as a Stackelberg game and used alternating direction method of multipliers (ADMM) to find the optimal solution. However, this work assumes that the entities involved reveal their private information, which may affect fairness in real-world scenarios. In addition, conventional methods such as ADMM require accurate network information to achieve optimal results and may have to re-solve the optimization problem again with the slightest change in traffic conditions, leading to huge computation overhead and poor convergence.

Recent advances in reinforcement learning (RL) has shown its ability to learn the stochastic policy of a dynamic environment without prior knowledge \cite{9239890}\cite{8657779}. 
Some authors proposed a deep RL (DRL) method for dynamic network management and resource allocation in IIoT network \cite{9239890}. 
Yao \textit{et al.} \cite{8657779} studied the resource management problem between a cloud provider and miners as a non-cooperative Stackelberg game, designing a multi-agent RL (MARL) algorithm to obtain the Nash equilibrium. However, fully non-cooperative games empower egoistic market players to maximize their own utilities even to the extent of degrading others’ utilities, without contributing to the overall system benefits.

Based on the above-mentioned limitations, this paper seeks to integrate a hybrid of cooperative and Stackelberg games, and MADRL to design a comprehensive economic framework for incentivized resource trading among sliced tenants in IIoT. We model the business interactions among the seller and buyer tenants as a hybrid cooperative Stackelberg game. 
Specifically, we 
formulate a coalition formation game where the buyer tenants choose to join coalitions with a higher chance of obtaining resource, as opposed to striving for resource as an individual entity. 
Then, a two-stage multi-leader multi-follower (MLMF) Stackelberg game is formulated between the seller tenants and buyer tenant coalitions, where the seller tenants as leaders set their unit price first, and the buyer tenant coalitions as followers determine their purchasing amount. 
We achieve a Stackelberg equilibrium (SE) for the formulated game by developing an MADRL method to make flexible pricing and purchasing decisions, without prior knowledge. Our main contributions are summarized as follows:
\begin{itemize}
	\item 	We design a novel strategic business framework for resource trading between virtualized tenants for NS in IIoT network. 
	\item   We formulate the interactive behavior of the market entities as a cooperative Stackelberg game based on their pricing and purchasing strategies for incentive maximization. 
	In the formulated game, buyer tenants (followers) strive to form coalitions in order to combat the pricing decisions of the seller tenants (leaders), provided they have the highest summed reputation score. Members would join the coalition if and only if they can gain more benefits than they could earn individually.
	\item   Considering the high-dimensional strategy space of the game players, we integrate the Stackelberg game model and multi-agent deep deterministic policy gradient (MADDPG) to propose a novel \textit{cooperative Stackelberg MADDPG} algorithm that ensures quick decision making for joint optimal pricing and purchasing strategies.
	
\end{itemize} 

The rest of the paper is organized as follows: Section II presents the system model, and Section III presents the joint intelligent pricing and purchasing-based resource management 
problem formulation. Simulation results and analysis are discussed in Section IV, and Section V concludes this work. 

\section{System Model}
We consider a single-cell time-synchronized OFDMA IIoT network where user equipment (UEs) of varying QoS requirements coexist. The system architecture as depicted in Fig. 1, consists of an MNO, multiple tenants, and a network controller. The physical network owned by the MNO is virtualized into logical networks, and assigned to different tenants who offer differentiated services to their respective UEs. We define three (3) tenants that form an application-specific smart manufacturing service based on 5G use cases as; \textit{smart factory} that provides enhanced mobile broadband (eMBB) service, \textit{smart energy} that provides ultra reliable low latency communication (uRLLC) service, and \textit{smart logistics} that provides massive machine type communication (mMTC) service \cite{9732420}. The network controller is in charge of network orchestration and management, by allocating resources to the tenants upon request.
\subsection{Business Model}
We consider a two-tier business model consisting of an MNO and multiple tenants in the resource trading market. The substrate infrastructure and resource are owned by MNO $i$, who leases them to a set of $j\in\mathcal{J}=\{1 ,2, ....,J \}$ tenants, to be subleased to their respective UEs. After the MNO leases resources to the tenants, changes in network conditions such as fluctuating network traffic behavior, compel the tenants to readjust their resource pools to suit their optimization goals. In this case, the tenants with extra resource to spare are incentivized to sublease a portion of their unused resource to the tenants in need of extra resources, for revenue in return. 
In this sequel, we refer to tenants who sublease their resource to other tenants as \textit{\textquote{sellers}} and tenants who demand extra resources as \textit{\textquote{buyers}}. Therefore, $\mathcal{J}$ comprises a set of $m \in \mathcal{M} = \{1 ,2, ....,M \}$ seller tenants and a set of $n \in \mathcal{N} = \{1 ,2, ....,N \}$ buyer tenants i.e. $\mathcal{M} \cup \mathcal{N}=\mathcal{J}$. Members of $\mathcal{N}$ can form coalitions as $\{z_{n}\} \in \mathcal{Z} = \{\{z_{1}\}, \{z_{2}\},...., \{z_{N}\} \}$ to stand a chance of combating the pricing strategies of the seller tenants. An $\textit{m-th}$ seller tenant will be willing to sell resource to buyer tenant coalition $\{z_{n}\}$ if the said coalition achieves the highest summed reputation score $\Omega_{\{z_{n}\}}$ of subleasing resource to other buyers in need. In this work, we refer to network resource as bandwidth $\mathcal{B}$, with granular units of $w \in \mathcal{W} = \{1 ,2, ....W \}$ physical resource blocks (PRBs). Each seller tenant $m$ has a maximum allowable PRBs $w_{m}^{max}$ that it can sell to a buyer tenant coalition $\{z_{n}\}$ at timeslot $t$, provided its required PRBs $w_{m}^{req}$ has been met, 
i.e., $0 \leq w_{\{z_{n}\}}(t) \leq w_{m}^{max}$.
\captionsetup[figure]{labelformat={default},labelsep=period,name={Fig.}}
\begin{figure} 
	\centering
	\includegraphics[width=0.45\textwidth]{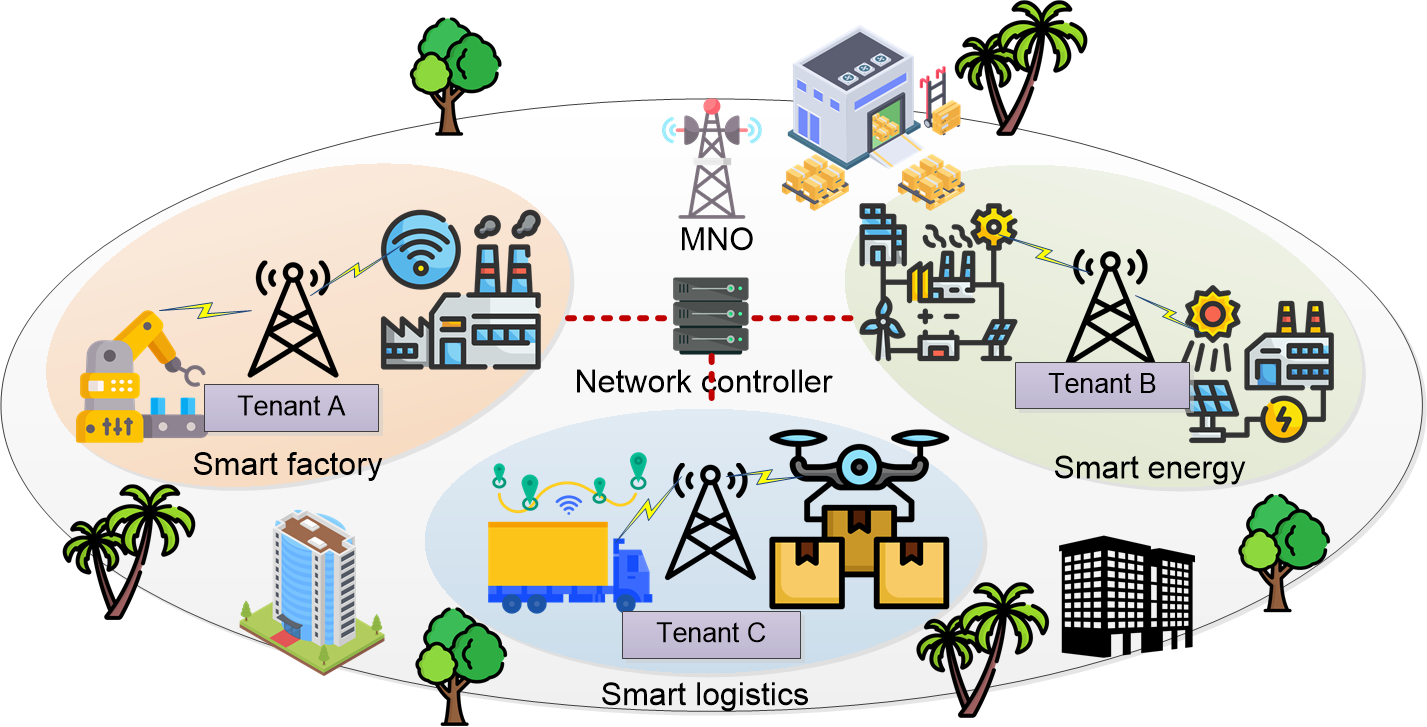}
	\caption{System architecture.}
	\label{framework}
\end{figure}

\subsection{Network Model}
Each tenant $j \in \mathcal{J}$ serves a set of $k \in \mathcal{K} = \{1 ,2, ....K \}$ UEs, thus $\mathcal{K}=\cup_{j\in\mathcal{J}}\mathcal{K}_{j}$. 
Each PRB has a bandwidth of $b_{w}$ Hz at every timeslot $t$, i.e., ${\sum_{w=1}^{W}b_{w}=\mathcal{B}}$. We assume that contiguous PRBs are allocated to each UE and that the channel gains of the PRBs are identically and independently distributed (i.i.d) \cite{8368322}. Let $x_{w,k}$ represent the binary PRB assignment indicator where $x_{w,k}=1$ means the PRB $w\in\mathcal{W}$ is assigned to UE $k\in\mathcal{K}$, and $x_{w,k}=0$ means otherwise. The maximum transmit power of the base station owned by MNO $i$ is $P_{i}^{max}$.

From the Shannon capacity formula \cite{9070169}, the achievable instantaneous data rate of a rate-sensitive (eMBB service) UE $k$ in tenant $j$ is a function of the PRBs allocated to it, and can be calculated as; 
\begin{equation}r_{k,j}=b_{w} \cdot \mathcal{B} \cdot \log_{2}(1 + \phi_{k,j}), \end{equation}
where $\phi_{k,j}$ denotes the signal-to-interference-plus-noise ratio (SINR). Then, the average achievable data rate of tenant $j$ is expressed as; 
\begin{equation} r_{j}={\sum_{k=1}^{K}r_{k,j}}.
\end{equation} 
The value of $r_{j}$ should meet the minimum data rate requirement ${r}_{j}^{min}$ of tenant $j$, i.e., $r_{j} \geq {r}_{j}^{min}$.

Next, we define the delay-QoS characteristics of a delay-sensitive (uRLLC service) UE based on the incoming service request at timeslot $t$. We assume the packet arrival of each UE $k$ in tenant $j$ follows a Poisson process with an average rate of $\lambda_{k,j}$ \cite{8792072}. Based on M/M/1 queuing theory, the achievable instantaneous delay of a packet is;
 \begin{equation}\tau_{k,j}=\frac{1}{{r_{k,j}}- \lambda_{k,j}}. \end{equation}
where $r_{k,j}$ and $\lambda_{k,j}$ are the achievable instantaneous data rate and packet arriving rate, respectively of UE $k$ in tenant $j$. The average delay of tenant $j$ is expressed as; \begin{equation}\tau_{j}={\sum_{k=1}^{K}\tau_{k,j}}.
\end{equation}
Similarly, the value of $\tau_{j}$ should meet the maximum delay requirement ${\tau}_{j}^{max}$ of tenant $j$, i.e., $\tau_{j} \leq {\tau}_{j}^{max}$. 

An mMTC service-based UE generally requires low data rate and is very tolerable to delay. Therefore, we assume a minimum of one assignable PRB should guarantee its data rate and delay requirements \cite{9220787}, i.e. ${\sum_{k\in \mathcal{K}}x_{w,k}\geq1}$.

\subsection{Utility Model}
To create a good impression about its services, a seller tenant cares about the QoS satisfaction of the buyer tenant coalitions. 
We model the QoS satisfaction on data rate ${r}_{j}$ of tenant $j$ as a sigmoid function \cite{4908302}, and is expressed as;
 \begin{equation} \xi  \left({r}_{j} \right) =\frac{1}{1+e^{- \eta  \left( {r}_{j}-{r}_{j}^{min} \right) }}, \end{equation}
where $\eta$ is used to adjust the utility curve around ${r}_{j}^{min}$.
and ${r}_{j}^{min}$ is the minimum data rate requirement of tenant $j$. 
Likewise, the QoS satisfaction on delay ${\tau}_{j}$ of tenant $j$ can be expressed as;
 \begin{equation} \xi  \left( {\tau}_{j} \right) =\frac{1}{1+e^{- \eta  \left( {\tau}_{j}^{max}- {\tau}_{j} \right) }}, \end{equation}
where ${\tau}_{j}^{max}$ is the maximum tolerant delay requirement of tenant $j$.

1) \textit{Utility Function of Seller Tenant:} Considering the unit price $\delta_m$(\$/Hz) of the $\textit{m-th}$ seller tenant and the PRB purchasing amount $w_{z_{n}}$ of a buyer tenant coalition ${z_{n}}$, the seller tenant's utility $\mathcal{U}_{m}$ is given by;
\begin{equation}\mathcal{U}_{m} = (\delta_{m} \cdot w_{z_{n}}(\cdot)) - (\delta_{i} \cdot w_{z_{n}}(\cdot)),   
\end{equation}
where $\mathcal{R}_{m}= (\delta_{m} \cdot w_{z_{n}}(\cdot))$ is the revenue seller tenant $m$ receives from selling PRBs to buyer tenant coalition ${z_{n}}$, and $\mathcal{C}_{m}=(\delta_{i} \cdot w_{z_{n}}(\cdot))$ is the cost involved in leasing the said PRBs from MNO $i$. We substitute $(\cdot)$ with customized versions of $\xi  \left({r}_{j} \right)$ or $\xi  \left( {\tau}_{j} \right)$ in (5) and (6) respectively, depending on the QoS requirement and the role of the tenant in PRB trading. We note that the purchasing amount of a buyer depends on its QoS demand and the selling price of a seller's PRB.

2) \textit{Utility Function of Buyer Tenant Coalition:} A group of buyer tenants may prefer to form a coalition $z_{n}$ with the aim of gathering the highest aggregated reputation $\Omega_{z_{n}}$ to be selected by the seller tenant as the winning coalition. However, this coalition formation comes with the cost of extra signaling among the coalition members to exchange essential information such as $\Omega_{n}$. 
We define the utility ${\mathcal{U}_{z_{n}}}$ of buyer tenant coalition $z_{n}$ as; 
\begin{equation}
{\mathcal{U}_{z_{n}}}= v(z_{n}) - (\mathcal{C}_{z_{n}} + \mathcal{C}_{sig}),
\end{equation}
where $v(z_{n})={\sum_{n \in {z_{n}}}} (\Omega_{n} \cdot w_{z_{n}}(\cdot))$ is the coalition value with signaling cost complexity $\mathcal{O}{|\mathcal{C}_{sig}|}^2$, $w_{z_{n}}(\cdot)$ is the purchasing amount based on QoS, and $\mathcal{C}_{z_{n}}=(\delta_{m} \cdot w_{z_{n}}(\cdot))$ is the cost of obtaining PRBs from seller tenant $m$.

\section{Problem Formulation}
\subsection{Coalition Formation for Buyer Tenants}
We formulate $\mathcal{N}$ buyer tenants' quest to form cooperative groups to stand a chance of negotiating with $\mathcal{M}$ seller tenants as a coalition formation game. Coalitions are formed to obtain the summed reputation of the buyer tenant coalition members, which is used by a seller tenant to determine the winning coalition. We define the coalition formation game as $\mathcal{G}=(\mathcal{N}, v)$, where $\mathcal{N}$ is the set of buyer tenants and $v$ is the coalition value that quantifies the worth of the coalition. It is noteworthy that any coalition ${z_{n}} \subseteq \mathcal{N}$ implies an agreement among members of ${z_{n}}$ to strive for PRBs as a single buyer. Based on $\mathcal{G}$, we present some basic definitions in the coalition formation game as follows:

1) \textit{Characteristic Form:} The value of coalition ${z_{n}}$ depends solely on the members of the coalition, with no dependence on how the players in $\mathcal{N}\setminus {z_{n}}$ are structured \cite{5230848}.

2) \textit{Transferable Utility (TU):} Coalitions formed with TU means that the total utility represented by a real number $\mathbb{R}$ can be divided in any manner among the coalition members.

\textbf{\textit{Definition 1 (Characteristic Form with TU):}} The value of the game $\mathcal{G}$ in characteristic form with TU is the function over $\mathbb{R}$ defined as $\mathcal{G}=v:2^{N} \rightarrow \mathbb{R}$, and the amount of utility that a player $n \in {z_{n}}$ receives from the division of $v(z_{n})$ constitutes its payoff $u_{n} \in \mathbb{R}^{|{z_{n}}|}, n \in {z_{n}}$.  

\textbf{\textit{Definition 2 (Stable Coalition Partition):}} For coalition partition ${z_{n}}$, no buyer tenant $n$ can improve its utility by switching to another coalition, i.e., ${\mathcal{U}_{z_{n}}}({w^{\ast}_{z_{n}}}, {w^{\ast}_{-{z_{n}}}}) \geq {\mathcal{U}_{z_{n}}}({w_{{z_{n}}}}, {w^{\ast}_{-{z_{n}}}}), \forall {z_{n}} \in \mathcal{N}, {w_{{z_{n}}}} \neq {w_{-{z_{n}}}}$.

The coalition formation process is explained below:

1) Initially, all the buyer tenants in the network are disjoint as in the set $\mathcal{N} = \{\{1\}, \{2\},...., \{N\} \}$. 

2) To form a strong force to combat the pricing strategy of a seller tenant, a group of buyer tenants form a coalition ${z_{n}}$ to aggregate a reputation score $\Omega_{z_{n}}$. 
We assume that two coalitions ${z_{1}}$ and ${z_{2}}$ can merge if the following constraint is satisfied: $v({z_{1}} \cup {z_{2}}) > v({z_{1}}) + v({z_{2}})$.

3) After forming coalitions, the seller tenant observes the coalition structures and selects the winning coalition as the one with the highest $\Omega_{z_{n}}$.

With the reputation-based cooperation, the coalition members obtain a portion of the PRBs in a fair manner (given their individual contributions) using $v(z_{n})$.  

\subsection{Stackelberg Game Formulation}
After the winning buyer tenant coalition is selected by seller tenant $m$, the two entities form a new game model, i.e. a Stackelberg game model. With the Stackleberg game, the buyer tenant coalition is able to negotiate and renegotiate the unit price offered by the seller tenant. We model the PRB trading interactions between the seller tenants and buyer tenant coalitions in the IIoT network as a two-stage Stackelberg game, where the seller tenants are the leaders and the buyer tenant coalitions are the followers. Specifically, the seller tenant $m$ first sets its unit price $\delta_{m}$ and then the buyer tenant coalition $z_{n}$ responds by deciding its purchasing amount $w_{z_{n}}$. It is noteworthy that $w_{z_{n}}$ is the aggregated expected purchasing amount of the buyers that form the coalition.
Each entity in the trading framework selects its strategy to maximize its own utility given the other entity's strategy. 
Both leaders and followers can adjust their strategies to maximize their respective utilities. We transform the two-stage game model into an equivalent PRB optimization problem as follows: 

1) \textit{Stage I: Leader's Price Imposition:} An $\textit{m-th}$ seller tenant sets its pricing strategy to maximize its utility $\mathcal{U}_{m}$ in (7), with the following optimization problem;
\begin{equation}
{\mathop{max}_{{\delta_{m} \geq 0}\\ }\mathcal{U}_{m}(w_{z_{n}}, \delta_{m})},\end{equation}
\begin{equation} 
\textit{s.t:} \quad  {\sum^{ {z_{N}}}_{{z_{n}} {=1}}{w_{{z_{n}}}}\leq w_{m}^{max}}, \end{equation} 
where $\mathcal{U}_{m}(w_{{z_{n}}}, \delta_{m})$ denotes the utility of the $\textit{m-th}$ seller tenant, $\delta_{m}$ and $w_{{z_{n}}}$ are the unit price and purchasing amount vectors with $[{\delta_{{m}_{1}}}, {\delta_{{m}_{2}}},...,{\delta_{M}}]^{T}$ and $[w_{{z_{1}}}, w_{{z_{2}}},..., w_{{z_{N}}}]^{T}$, respectively. Constraint (10) ensures that the purchasing amount of the buyer tenant coalition cannot exceed the maximum allowable PRBs that can be sold by the seller tenant.

2) \textit{Stage II: Follower's Purchasing Amount Response:} Considering $\delta_{m}$, the buyer tenant coalition determines its purchasing strategy to maximize its utility in (8), with the following optimization problem;
\begin{equation}
{\mathop{max}_{{w_{z_{n}} \geq 0}\\ }\mathcal{U}_{z_{n}}(w_{{z_{n}}}, \delta_{m})}.\end{equation}
We use (9) and (11) to form the Stackelberg game with the objective of finding an SE, where neither of the entities in the game has an incentive to deviate.

\textbf{\textit{Definition 3 (Stackelberg Equilibrium):}} Given the optimal unit price and purchasing amount of seller tenant $m$ and buyer tenant coalition $z_{n}$ as $\delta^{\ast}_{m}$ and $w^{\ast}_{{z_{n}}}$ respectively, the SE is $(\delta^{\ast}=\{\delta_{m}\}_{m \in \mathcal{M}}$, $w^{\ast}=\{w_{z_{n}}\}_{z_{n} \in \mathcal{N}})$  , if

1) For any buyer tenant coalition $z_{n} \in \mathcal{N}$, given all seller tenants choose their optimal prices, buyer tenant coalition $z_{n}$ chooses its optimal purchasing amount $w^{\ast}_{z_{n}}$ to maximize its utility $\mathcal{U}_{z_{n}} (w^{\ast}_{z_{n}},\delta^{\ast})\geq \mathcal{U}_{z_{n}} (w_{z_{n}},\delta^{\ast} )\forall z_{n} \in \mathcal{N}.$

2) For any seller tenant $m \in \mathcal{M}$, given all buyer tenant coalitions choose their optimal purchasing amounts, seller $m$ chooses its optimal price $\delta^{\ast}_{m}$ to maximize its utility $\mathcal{U}_{m} (\delta^{\ast}, w^{\ast}_{z_{n}})\geq \mathcal{U}_{m} (\delta, w^{\ast}_{z_{n}})\forall m \in \mathcal{M}.$

To verify the existence and uniqueness of the SE, we take the second order derivatives of (7) and (8) \cite{8648014}\cite{9833469}. It is proven in literature that backward induction can be used to achieve SE for the formulated game. However, this method of finding SE requires full and accurate game information, which may affect the fairness of the game. Acquiring accurate game information by conventional means seem impractical since the buyer tenant coalitions and the seller tenants continue to negotiate and renegotiate at time intervals in order to achieve their respective optimal utilities. In contrast, DRL approach learns the optimal policy without prior knowledge. Therefore, we design a DRL-based method for obtaining joint optimal pricing and purchasing strategies for PRB optimization; hence, achieving the SE.

\subsection{MADRL-based Algorithm for Utility Optimization}

We transform the pricing and purchasing problem in the sliced IIoT network as a stochastic Markov decision process (MDP), and propose a solution based on DRL technique. 
The purpose of our DRL approach is to find optimal pricing and purchasing strategies of the seller tenants and buyer tenant coalitions that solves the Stackelberg game, with no prior knowledge. Each entity is assigned a learning agent that gathers network information from the environment as the conditions of trading keeps changing in real-time. 
Since the seller tenants and buyer tenant coalitions have different objectives in the game, an MADRL system is preferred to a single-agent DRL system. This is because a single agent only maximizes its own cumulative reward, while a multi-agent maximizes the cumulative reward of all agents to achieve their individual and common objectives. 
A detailed MDP formulation can be found in our prior work in \cite{9594082}.

From Markov property, the policy $\pi$ can be obtained by;
\begin{equation} 
\mathcal{V}^{\pi }\left(s\right) {=}\mathbb{E}_{\pi}\left\{r^t {+\ } \gamma \sum_{s^{ {t+1}}}{P\left(s^{ {t+1}}\mathrel{\left|\vphantom{s^{ {t+1}} s^t,a^t}\right.\kern-\nulldelimiterspace}s^t,a^t\right)\mathcal{V}^{\pi }\left(s^{ {t+1}}\right)}\right\},        
\end{equation}
where $r^t$ is the present reward, $\mathcal{V}^{\pi}(s)$ is the present utility, and $\mathcal{V}^{\pi}{(s^{t+1})}$ is the future utility. The state-value function for an optimal policy based on Bellman equation \cite{Sutton1998} is given as;
\begin{equation} 
\mathcal{V}^{{\pi }^{ {*}}}\left(s\right) {=}{arg \mathop{max}_{{a^t}\in {A}}\\ }\left\{\mathcal{V}^{\pi }\left(s\right)
\right\}.       
\end{equation}
We begin to define the components of the MDP tuple as follows:

\textit{State(s)}: Since the seller tenant sets its unit price first, it observes the purchasing strategy of the buyer tenant coalition at the previous timeslot $t-1$. Simultaneously, the buyer tenant coalition observes the current unit price of the seller tenant to decide its purchasing amount. Therefore, the states of seller tenant $m$ and buyer tenant coalition $z_{n}$ at timeslot $t$ are given by $s^{t}_{m}=\{w_{z_{n}}^{t-1}\}_{z_{n} \in \mathcal{N}}$ and $s^{t}_{z_{n}}=\{\delta^{t}_{m}\}_{m \in \mathcal{M}}$, respectively.

\textit{Action(a)}: 
At timeslot $t$, the seller tenant $m$ sets its unit price from the set of possible actions as $a^{t}_{m} \in \mathcal{A}_{m}$, and then the buyer tenant coalition $z_{n}$ decides its purchasing amount from the set of possible actions as $a^{t}_{z_{n}} \in \mathcal{A}_{z_{n}}$. For simplicity, we assume that $m$ cannot sell more than half of its PRBs to $z_{n}$, i.e., we define $\mathcal{A}_{m}$ and $\mathcal{A}_{z_{n}}$ as $\mathcal{A}_{m}=\{1, 2,...., 100\}$ and $\mathcal{A}_{z_{n}}=\{1, 2, ....50\}$.

\textit{Reward(r)}: To maximize the long-term utility of a seller tenant $m$ and buyer tenant coalition ${z_n}$, we define the immediate reward $r^{t}_{m}$ and $r^{t}_{z_{n}}$ based on their respective utility functions as $r^{t}_{m}=\mathcal{U}_{m}(w_{z_{n}}^{t-1}, \delta^{t}_{m})$ and $r^{t}_{z_{n}}=\mathcal{U}_{z_{n}}(w^{t}_{z{n}}, \delta^{t}_{m})$, respectively. The system utility is therefore $r={\sum_{m=1}^{M}}{\sum_{z_{n}=1}^{z_{N}}}(r^{t}_{m}+r^{t}_{z_{n}})$.

We deploy an MADDPG algorithm named \textit{cooperative Stackelberg MADDPG}, to achieve the SE of the formulated game. The DDPG architecture adopts an actor-critic approach that combines the gains of policy-based and value-based methods. By policy function, the actor generates an action given a state. The critic produces an action-value function and uses a loss function to criticize the actor's performance. Then, the actor uses DPG to approximate policies with the critic's output. DPG directly generates deterministic behavior policy, and avoids frequent action sampling. The critic updates the action-value function using gradient descent method \cite{9322481}. 

The actor chooses an action $a^{t}$ based on current state $s^{t}$ and current policy $\pi$ as;
\begin{equation} 
a^{t}=\pi(s^{t}, \theta^{\pi}).        
\end{equation}
Based on the Bellman equation, the critic network calculates the target Q-value as;
\begin{equation} 
y^{t}{=}r^{t} + \gamma \cdot Q{'}(s{'}, \pi{'}, (s{'}{|}\theta^{\pi}{'}), \theta^{Q}{'})).        
\end{equation}
Let $\pi_{m}$ and $\pi_{z_{n}}$ be the set of policies for seller tenant $m$ and buyer tenant coalition $z_{n}$, respectively where ${\pi}_{m}$=$\{\pi_{1},....,\pi_{M}\}$, and ${\pi}_{z_{n}}$=$\{\pi_{1},....,\pi_{z_{N}}\}$.

At \textit{Stage I}, the critic network can be updated by minimizing the loss function as;
\begin{equation} 
\begin{aligned}
\mathcal{L}_{m}(Q_{m})=\mathbb{E}_{(s_{m}, a_{m}, r_{m}, s{'}_{m})\sim \mathcal{D}_{m}} [(y_{m}-Q_{m}(s, a_{m};\theta_{m}))^2] \\ 
y_{m}=r_{m} + \eta \cdot StackelbergQ_{m}(s{'}), \\ 
\end{aligned}       
\end{equation}
where $\textit{Stackelberg}Q_{m}(s{'})=\mathop{max}_{a{'}} Q_{m}(s{'}, a{'}_{m}, \theta_{m})$ is the SE reward under state $s{'}$.
\begin{center}
\begin{algorithm} [!t] 
\caption{Cooperative Stackelberg MADDPG Algorithm}
\begin{algorithmic}[1]
	\State \textbf{Randomly initialize:} Actor and critic evaluation networks with random weights $\theta^{\pi}$ and $\theta^{Q}$, respectively
	\State \textbf{Initialize:} Actor and critic target networks with weights $\theta^{\pi}{'} \gets \theta^{\pi}$ and $\theta^{Q}{'} \gets \theta^{Q}$, respectively
	\State \textbf{Initialize:} Replay memory $D$ and mini-batch $D^{'}$
	\For {each iteration}
	\State Set up the simulation environment
	\For {each decision step $t$, }
	\For {each agent} 
	\State Observe state $s^{t}$
	\State Design coalition formation game for $\mathcal{N}$ buyers
	\State Stackelberg game for PRB trading with (9),(11)
	\State Select action $a^{t}$ for exploration based on (14)  
	\State Perform $a^{t}$, compute $r^t$ and $s^{t+1}$
	\State Update resource pool at BS-level			
	\State Store experience ($s^t$, $a^t$, $r^t$, $s^{t+1}$) in $D$
	\State Sample mini-batch of transitions from $D$
	\State Compute target value $y^t$ using (15)
	\State Update critic network by $\mathcal{L}(Q)$ using (16)
	\State Update actor network by $\bigtriangledown_{\theta^{\pi}}J(\pi)$ using (17) 
	\State Update target networks by soft update via (18)
	\EndFor 
	\EndFor
	\EndFor
\end{algorithmic}
\end{algorithm}
\end{center}
The policy gradient of the DPG objective function with respect to $\theta^{\pi_{m}}$ is given by;
\begin{equation} 
\begin{aligned}
\bigtriangledown_{\theta^{\pi_{m}}}J(\pi_{m}){=}\mathbb{E}_{s, a_{m}\sim \mathcal{D}_{m}}
[\bigtriangledown_{\theta^{\pi_{m}}}\pi_{m}(a_{m}, s_{m}) \bigtriangledown_{a_{m}} Q_{m}(s, a_{m}, \\ \theta^{Q}{|}a_{m}=\pi_{m}(s_{m}))]. \\
\end{aligned}        
\end{equation}
Finally, we update the target network of $m$, using soft update;
\begin{equation} 
\begin{aligned}
\theta^{\pi_{m}{'}} \gets \tau\theta^{\pi_{m}} + (1-\tau)\theta^{\pi_{m}{'}},
\theta^{Q_{m}{'}} \gets \tau\theta^{Q_{m}} + (1-\tau)\theta^{Q_{m}{'}}, \\
\end{aligned}    
\end{equation}
where $\tau$ denotes the learning rate. \textit{Stage II} follows a similar formulation to compute  $\bigtriangledown_{\theta^{\pi_{z_{n}}}}J(\pi_{z_{n}})$, $\mathcal{L}_{z_{n}}(Q_{z_{n}})$, $\theta^{\pi_{z_{n}}{'}}$, $\theta^{Q_{z_{n}}{'}}$ for the buyer tenant coalition $z_{n}$.

A detailed cooperative Stackleberg MADDPG algorithm is presented in \textbf{Algorithm 1}. The computational complexity of the MADDPG algorithm is expressed as $\mathcal{O}(\mathcal{G} \times |\mathcal{S}|\times|\mathcal{A}|)$, where $\mathcal{G}$ denotes the total number of agents, $\mathcal{S}$ denotes the state set, and $\mathcal{A}$ denotes the action set. Let the number of hidden layers be $H$ and the dimension of the output be $L$. The complexity of each actor and critic network is $\mathcal{O}({|L|}^{2} {H})$.

\section{Performance Evaluation}
\begingroup
\begin{table} [b]
\centering
\caption{Simulation Parameters}
\smallskip\noindent
\resizebox{\linewidth}{!}{%
\label{Simpara}
\setlength{\tabcolsep}{3pt}
\begin{tabular}{|l|l|}
	\hline
	\textbf{Parameters and Units} & 
	\textbf{Values} \\\hline 
	Number of tenants, $\mathcal{J}$  &
	$5$  \\ \hline
	Number of users, $\mathcal{K}$  & 50 in each tenant \\ \hline
	System bandwidth, $\mathcal{B}$ & 20 MHz \\ \hline
	Number of PRBs, $\mathcal{W}$ & 100  \\ \hline
	Transmit power of BS, $P_i$ &  30 dBm  \\ \hline
	Network coverage area & 500 m $\times$ 500 m \\ \hline
	Noise power density, $\theta^{2}$ & -174 dBm/Hz \\ \hline
	User distribution & Uniform \\ \hline
	Tenant minimum data rate (${r}^{min}$) &  [Tenant 1-2=500, Tenant 3-4=10, Tenant 5=15] kbps \\ \hline
	Tenant maximum delay (${\tau}^{max}$) & [Tenant 1-2=100, Tenant 3-4=10, Tenant 5=100] ms \\ \hline
	Number of hidden layers(actor and critic)  & 2 (128 neurons in each) \\ \hline
	Number of iterations    & 2500 \\ \hline 
	Discount factor, $\gamma_a$, $\gamma_c$  & 0.9 \\ \hline
	Replay memory size, $D$ & $10^5$ \\ \hline
	Mini batch size, $D^{'}$ & $128$  \\ \hline
	Learning rate, $\tau_a$,  $\tau_c$    &  0.001 \\ \hline
\end{tabular}}
\label{tab1}
\end{table}
\endgroup
In this section, we evaluate the performance of our proposed \textit{cooperative Stackelberg MADDPG} algorithm via simulation results and analysis. All simulations are performed in a Python 3.8 enviroment with TensorFlow 2.0, running on a core i7 server, 2.8GHz Intel Xeon CPU, and 16GB RAM. We consider a single-cell network of 500m $\times$ 500m BS coverage area, with a BS transmit power budget  and noise spectral density set to 30dBm and -174 dBm, respectively. The system bandwidth is set at 20MHz with 100 PRBs. We define two eMBB tenants, two URLLC tenants, and one mMTC tenant, with 50 users distributed in each. We assume a log-normal distribution for shadow fading and adopt the following path loss (PL) model: $PL (dB) = 20\log_{10}(d)+20\log_{10}(f)–27.55$, where $d$ and $f$ represent distance (in meters) and frequency (in MHz), respectively. 
At each run, the coalition with the highest reputation $\Omega_{z_{n}}$ is selected as the winning coalition. We define $\mathcal{A}_{m}$ and $\mathcal{A}_{z_{n}}$ as $\mathcal{A}_{m}=\{1, 2,...., 100\}$ and $\mathcal{A}_{z_{n}}=\{1, 2, ....50\}$, respectively. Quantitatively, the price of one PRB is in the range $\delta=[1.0,..., 2.0]\$/Hz$. 

For the MADDPG and DDPG algorithms, we set the size of replay memory, minibath size, and discount factor to $10^5$, 128, and 0.9 respectively. Each of the MADDPG and DDPG models consists of two fully-connected feed-forward neural networks for each actor and critic, with 128 neurons in each network. All parameters of the learning are derived from parameter tuning. We utilize $ReLU$ activation function for the hidden layers and $tanh$ for the output layer. All simulation results are averaged over a number of random independent runs. To optimize the loss, we adopt the \textit{AdamOptimizer}. Simulation parameters are summarized in Table I.
\begin{figure} 
	\centering
	\includegraphics[width=0.30\textwidth]{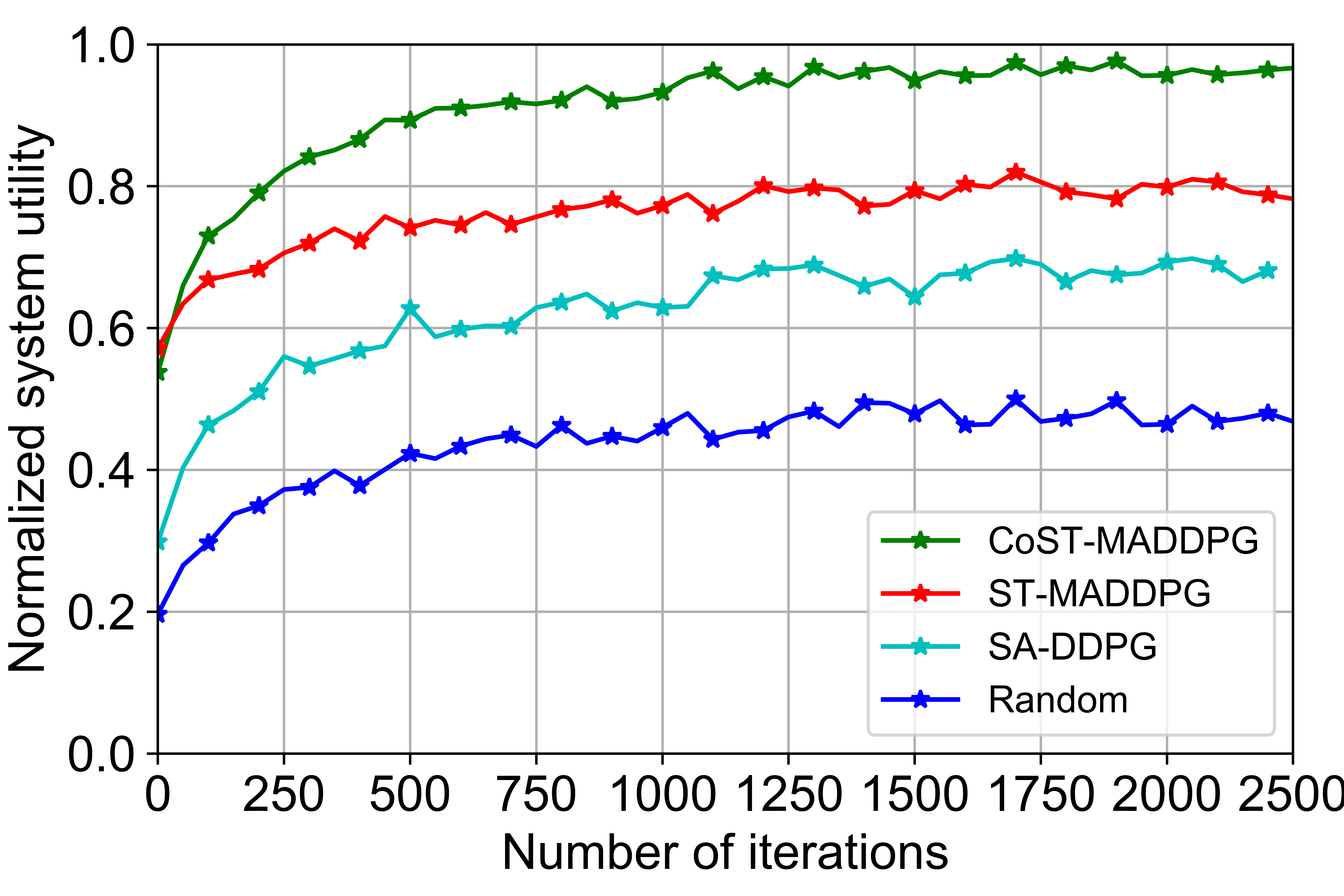}
	\caption{Convergence analysis.}
	\label{framework}
\end{figure}

\subsection{Convergence Analysis}
In this simulation, we verify the convergence performance of our proposed cooperative Stackelberg MADDPG (CoST-MADDPG) algorithm, with Stackelberg MADDPG (ST-MADDPG) \cite{9444840}, single-agent DDPG (SA-DDPG) \cite{9322481}, and Random algorithm (Random) as baselines. We run the simulation for 2500 iterations and the results are averaged over every 250 iterations for performance comparison. 
Fig. 2 shows the convergence on normalized system utility with increasing number of iterations, for the four algorithms. 
From Fig. 2, we observe that all the four algorithms achieve convergence with increasing number of iterations. Particularly, the proposed CoST-MADDPG algorithm achieves the fastest convergence and highest normalized system utility at about 500 iterations and 0.95, respectively. The ST-MADDPG algorithm achieves convergence at 500 iterations, but with system utility of about 0.80. The reason for this trend is that the proposed CoST-MADDPG takes advantage of coalition formation of buyers to enhance fairness in utility optimization of both sellers and buyers, which increases overall system utility. 
Among the learning methods, SA-DDPG achieves the worst results because it deploys a single agent, who maximizes its own reward. 
Of the four algorithms, Random algorithm achieves the worst convergence with the reason being that it selects pricing and purchasing actions with random probability. We can conclude that the proposed CoST-MADDPG algorithm can best learn the optimal policy to maximize overall system utility, compared with the other baselines.

\subsection{Impact on Pricing and Purchasing Strategies}
In Fig. 3, we compare the performance of CoST-MADDPG algorithm with the baselines, in terms of their impact on the pricing and purchasing strategies of the sellers and buyers in the trading market. 
For the baseline algorithms, we consider a scenario where 2 seller tenants trade PRBs with 3 buyer tenants. For CoST-MADDPG, the buyer tenants form a 3-member buyer tenant coalition. Fig. 3(a) and 3(b) show the pricing and purchasing trends of the four algorithms against the bandwidth available for trading, respectively. From Fig. 3(a), we observe that as the amount of bandwidth increases, the bandwidth price decreases in all four algorithms. For instance, with about 5MHz bandwidth, the bandwidth price of CoST-MADDPG, ST-MADDPG, SA-DDPG, and Random are approximately 35, 50, 58, and 70, respectively. With 20MHz bandwidth, the bandwidth prices decrease to about 25, 35, 41, and 61, respectively. We observe this trend because with a small amount of bandwidth, the sellers set higher prices due to scarce resource. However, as the amount of bandwidth increases, the sellers have a large amount of goods to sell, so they lower their prices to stimulate consumption. From Fig. 3(b), we observe that as the amount of bandwidth increases, the purchasing amount of the buyers increases, with the proposed CoST-MADDPG algorithm achieving the highest purchasing amount followed by ST-MADDPG, SA-DDPG, and Random in that order. The reason for this trend is that, the prices are reduced to motivate the buyers to buy bandwidth. 
We can conclude that the proposed algorithm is able to best match the pricing and purchasing strategies of the sellers and buyers due to its ability to find the SE which gives the optimal pricing and purchasing decisions.
\begin{figure}
\centering
\begin{subfigure}{0.25\textwidth}
\includegraphics[width=\textwidth]{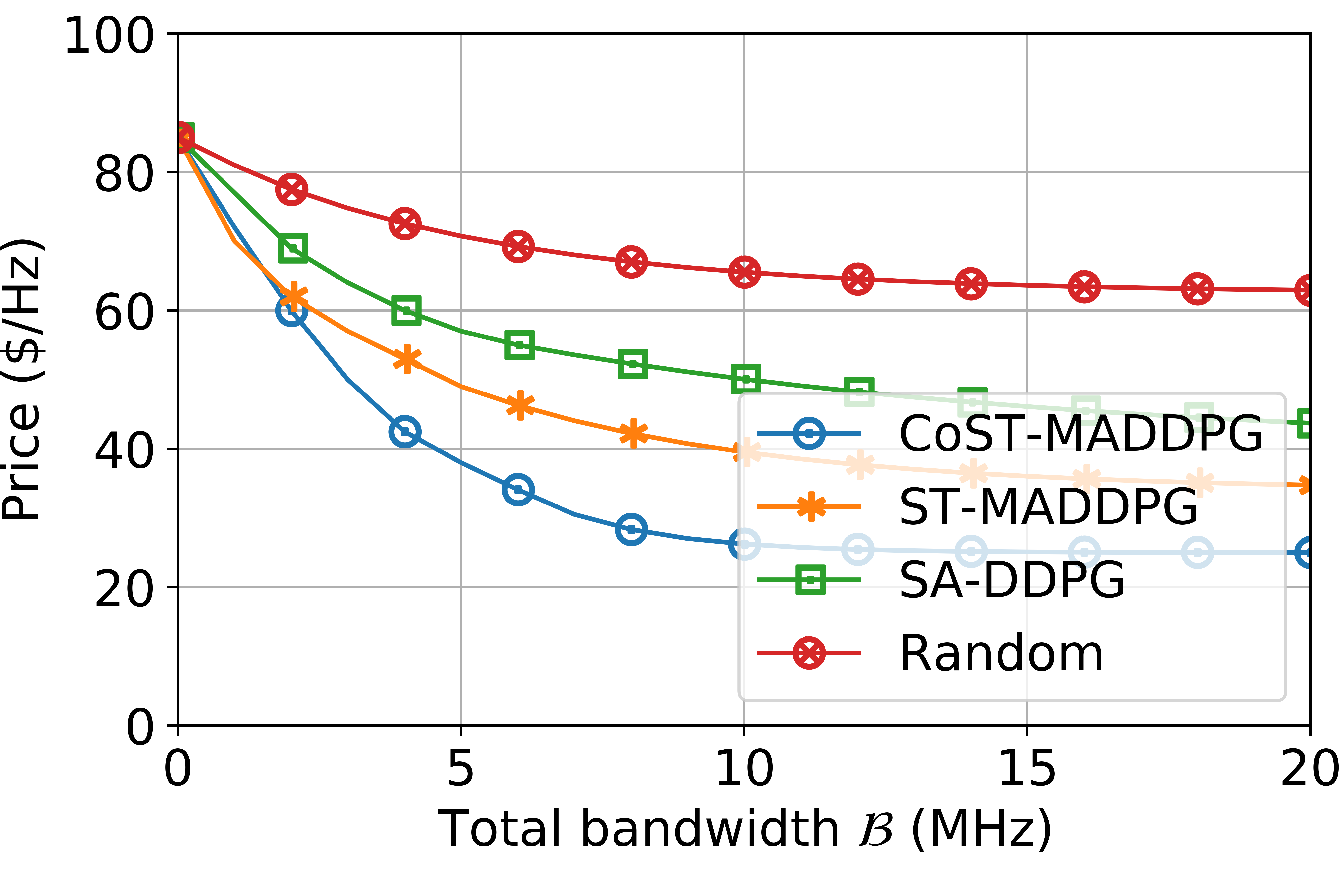}
\caption{seller tenants' price vs. $\mathcal{B}$.}
\label{fig:arf}
\end{subfigure}%
\begin{subfigure}{0.25\textwidth}
\includegraphics[width=\textwidth]{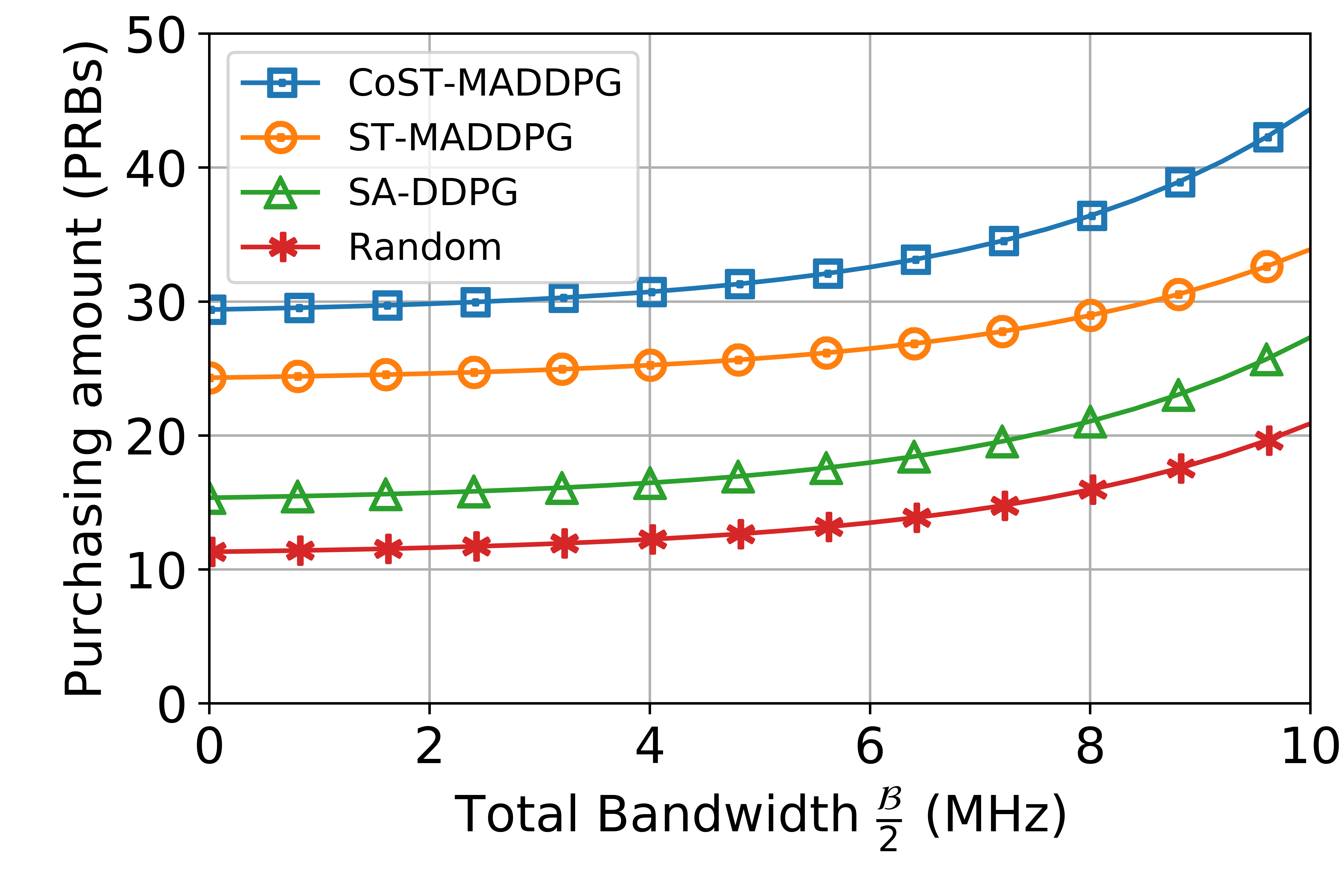}
\caption{Buyer's purchasing vs. $\mathcal{B}/2$.}
\label{fig:alb}
\end{subfigure}
\caption{Impact on pricing and purchasing strategies.}
\label{Timedelay}
\end{figure}

\subsection{Impact of Changing Number of Sellers and Buyers}
Since the CoST-MADDPG and ST-MADDPG algorithms achieve the best results in the previous subsections, in this simulation, we compare the performance of both algorithms in terms of changing number of entities (buyers and sellers) in the trading environment. We increase the number of seller tenants and buyer tenants to 5 each, in order to achieve more meaningful results. Fig. 4(a) and 4(b) show the performance of CoST-MADDPG and ST-MADDPG with increasing number of seller tenants and buyer tenants, respectively.

From Fig. 4(a), we observe that with only 1 seller tenant in the trading environment, both CoST-MADDPG and ST-MADDPG achieve very high utilities of about 92 and 91, respectively. This is so because with 1 seller tenant, there is a monopolistic market. That is, the buyer tenants or a coalition of them are forced to buy resources at a high unit price. However, as the number of seller tenants increases, competition among the seller tenants begin to exist and that the utility of the seller tenants under both algorithms decreases. For instance, with 5 seller tenants, the seller tenant utility under CoST-MADDPG is approximately 59 and that under ST-MADDPG is about 75. The proposed CoST-MADDPG algorithm achieves a lower seller tenant utility than ST-MADDPG because at this point, the buyer tenants may have formed coalitions to combat the pricing strategies of the tenants, forcing the sellers to further reduce their unit prices. With ST-MADDPG, the buyer tenants may act individually to negotiate prices with the seller tenants, which may not give them the chance to obtain lower prices.

From Fig. 4(b), we observe that with one buyer tenant, both CoST-MADDPG and ST-MADDPG achieve low buyer tenant utilities at about 15 and 12, respectively. The reason for this trend is that a one-member buyer coalition or one buyer tenant in the trading environment does not have much power to negotiate with a monopolistic seller or a number of sellers. As the number of buyer tenants increases, coalitions are formed in CoST-MADDPG to combat the pricing strategies of the seller tenants. This is evident with 5 buyer tenants, where CoST-MADDPG achieves a buyer tenant coalition utility of about 45. However, in ST-MADDPG, the individual buyer tenants act egoistically to negotiate and renegotiate the unit pricing with the seller tenants. Therefore, they are unable to achieve higher utility.

We can conclude that the proposed CoST-MADDPG algorithm is able to achieve acceptable levels for both seller tenants and buyer tenant coalitions, better than ST-MADDPG algorithm.
\begin{figure}
	\centering
	\begin{subfigure}{0.25\textwidth}
		\includegraphics[width=\textwidth]{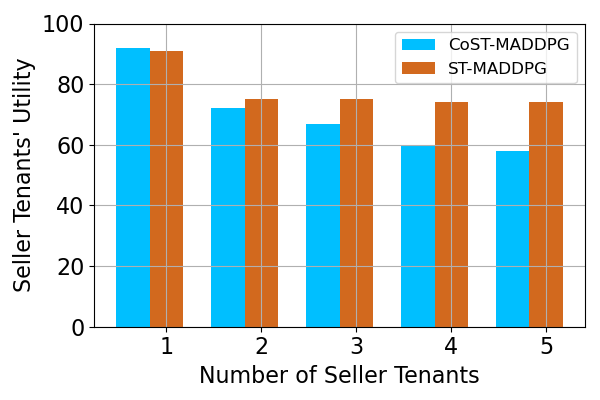}
		\caption{Changing no. of seller tenants.}
		\label{fig:arf}
	\end{subfigure}%
	\begin{subfigure}{0.25\textwidth}
		\includegraphics[width=\textwidth]{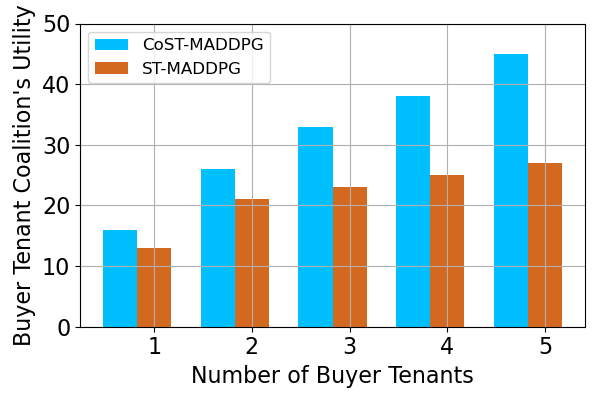}
		\caption{Changing no. of buyer tenants.}
		\label{fig:alb}
	\end{subfigure}
	\caption{Performance on changing no. of buyers and sellers.}
	\label{Timedelay}
\end{figure}

\section{Conclusion}
This paper designed a framework for the business interactions between seller tenants and buyer tenant coalitions in a sliced IIoT network. Particularly, we formulated the trading model as a cooperative Stackelberg game, where buyer tenants formed coalitions to combat seller tenants' price negotiations for resource trading. 
Then, a two-stage Stackelberg game was formulated to achieve optimal pricing and purchasing strategies for the seller tenants and buyer tenant coalitions, respectively. To achieve an SE, we developed a cooperative Stackelberg MADDPG method to learn the optimal strategies of the trading entities, without prior knowledge of the environment. Simulation results proved that the proposed method can converge to an optimal solution, and is able to best optimize the utilities of sellers and buyer tenant coalitions, compared with other benchmark algorithms.

\section*{Acknowledgments}
This work is supported in part by the Natural Science Foundation of China under Grant No.61806040 and Grant No. 61771098; in part by the fund from the Department of Science and Technology of Sichuan Province under Grant No. 2020YFQ0025; in part by the Natural Science Foundation of Guangdong Province under Grant No. 2021A1515011866; in part by the Social Foundation of Zhongshan Sci-Tech Institute under Grant 420S36; and in part by the fund from Intelligent Terminal Key Laboratory of Sichuan Province under Grant No. SCITLAB-1018.

%


\newcommand{\BIBdecl}{\setlength{\itemsep}{0.25 em}}
\bibliographystyle{IEEEtran}
\bibliography{ref}

\begin{thebibliography}{10}
\providecommand{\url}[1]{#1}
\csname url@samestyle\endcsname
\providecommand{\newblock}{\relax}
\providecommand{\bibinfo}[2]{#2}
\providecommand{\BIBentrySTDinterwordspacing}{\spaceskip=0pt\relax}
\providecommand{\BIBentryALTinterwordstretchfactor}{4}
\providecommand{\BIBentryALTinterwordspacing}{\spaceskip=\fontdimen2\font plus
\BIBentryALTinterwordstretchfactor\fontdimen3\font minus
  \fontdimen4\font\relax}
\providecommand{\BIBforeignlanguage}[2]{{%
\expandafter\ifx\csname l@#1\endcsname\relax
\typeout{** WARNING: IEEEtran.bst: No hyphenation pattern has been}%
\typeout{** loaded for the language `#1'. Using the pattern for}%
\typeout{** the default language instead.}%
\else
\language=\csname l@#1\endcsname
\fi
#2}}
\providecommand{\BIBdecl}{\relax}
\BIBdecl

\bibitem{9732420}
Y.~Wu, H.-N. Dai, H.~Wang, Z.~Xiong, and S.~Guo, ``A {S}urvey of {I}ntelligent
  {N}etwork {S}licing {M}anagement for {I}ndustrial {I}o{T}: {I}ntegrated
  {A}pproaches for {S}mart {T}ransportation, {S}mart {E}nergy, and {S}mart
  {F}actory,'' \emph{IEEE Communications Surveys Tutorials}, pp. 1--1, 2022.

\bibitem{9629333}
L.~Ji, S.~He, W.~Wu, C.~Gu, J.~Bi, and Z.~Shi, ``Dynamic {N}etwork {S}licing
  {O}rchestration for {R}emote {A}daptation and {C}onfiguration in {I}ndustrial
  {I}o{T},'' \emph{IEEE Transactions on Industrial Informatics}, vol.~18,
  no.~6, pp. 4297--4307, 2022.

\bibitem{9498080}
S.~Messaoud, A.~Bradai, S.~Dawaliby, and M.~Atri, ``Slicing {O}ptimization
  based on {M}achine {L}earning {T}ool for {I}ndustrial {I}o{T} 4.0,'' in
  \emph{2021 IEEE International Conference on Design Test of Integrated Micro
  Nano-Systems (DTS)}, 2021, pp. 1--5.

\bibitem{9686109}
T.~Umagiliya, S.~Wijethilaka, C.~De~Alwis, P.~Porambage, and M.~Liyanage,
  ``Network {S}licing {S}trategies for {S}mart {I}ndustry {A}pplications,'' in
  \emph{2021 IEEE Conference on Standards for Communications and Networking
  (CSCN)}, 2021, pp. 30--35.

\bibitem{8362975}
A.~E. Kalør, R.~Guillaume, J.~J. Nielsen, A.~Mueller, and P.~Popovski,
  ``Network {S}licing in {I}ndustry 4.0 {A}pplications: {A}bstraction {M}ethods
  and {E}nd-to-{E}nd {A}nalysis,'' \emph{IEEE Transactions on Industrial
  Informatics}, vol.~14, no.~12, pp. 5419--5427, 2018.

\bibitem{9646249}
C.~Chi, Y.~Wang, X.~Tong, M.~Siddula, and Z.~Cai, ``Game {T}heory in {I}nternet
  of {T}hings: {A} {S}urvey,'' \emph{IEEE Internet of Things Journal}, pp.
  1--1, 2021.

\bibitem{9763379}
Z.~Abou El~Houda, B.~Brik, A.~Ksentini, L.~Khoukhi, and M.~Guizani, ``When
  {F}ederated {L}earning {M}eets {G}ame {T}heory: {A} {C}ooperative {F}ramework
  to {S}ecure {I}{I}o{T} {A}pplications on {E}dge {C}omputing,'' \emph{IEEE
  Transactions on Industrial Informatics}, pp. 1--1, 2022.

\bibitem{9509582}
Y.~Jiang, Y.~Zhong, and X.~Ge, ``{I}{I}o{T} {D}ata {S}haring {B}ased on
  {B}lockchain: {A} {M}ultileader {M}ultifollower {S}tackelberg {G}ame
  {A}pproach,'' \emph{IEEE Internet of Things Journal}, vol.~9, no.~6, pp.
  4396--4410, 2022.

\bibitem{9239890}
S.~Messaoud, A.~Bradai, O.~B. Ahmed, P.~T.~A. Quang, M.~Atri, and M.~S.
  Hossain, ``Deep {F}ederated {Q}-{L}earning-{B}ased {N}etwork {S}licing for
  {I}ndustrial {I}o{T},'' \emph{IEEE Transactions on Industrial Informatics},
  vol.~17, no.~8, pp. 5572--5582, 2021.

\bibitem{8657779}
H.~Yao, T.~Mai, J.~Wang, Z.~Ji, C.~Jiang, and Y.~Qian, ``Resource {T}rading in
  {B}lockchain-{B}ased {I}ndustrial {I}nternet of {T}hings,'' \emph{IEEE
  Transactions on Industrial Informatics}, vol.~15, no.~6, pp. 3602--3609,
  2019.

\bibitem{8368322}
M.~Sinaie, D.~Wing Kwan~Ng, and E.~A. Jorswieck, ``Resource {A}llocation in
  {N}{O}{M}{A} {V}irtualized {W}ireless {N}etworks {U}nder {S}tatistical
  {D}elay {C}onstraints,'' \emph{IEEE Wireless Communications Letters}, vol.~7,
  no.~6, pp. 954--957, 2018.

\bibitem{9070169}
G.~Sun, G.~O. Boateng, D.~Ayepah-Mensah, G.~Liu, and J.~Wei, ``Autonomous
  {R}esource {S}licing for {V}irtualized {V}ehicular {N}etworks {W}ith {D}2{D}
  {C}ommunications {B}ased on {D}eep {R}einforcement {L}earning,'' \emph{IEEE
  Systems Journal}, vol.~14, no.~4, pp. 4694--4705, 2020.

\bibitem{8792072}
X.~Chen, Z.~Zhao, C.~Wu, M.~Bennis, H.~Liu, Y.~Ji, and H.~Zhang,
  ``Multi-{T}enant {C}ross-{S}lice {R}esource {O}rchestration: {A} {D}eep
  {R}einforcement {L}earning {A}pproach,'' \emph{IEEE Journal on Selected Areas
  in Communications}, vol.~37, no.~10, pp. 2377--2392, 2019.

\bibitem{9220787}
P.~K. Korrai, E.~Lagunas, A.~Bandi, S.~K. Sharma, and S.~Chatzinotas, ``Joint
  {P}ower and {R}esource {B}lock {A}llocation for {M}ixed-{N}umerology-{B}ased
  5{G} {D}ownlink {U}nder {I}mperfect {C}{S}{I},'' \emph{IEEE Open Journal of
  the Communications Society}, vol.~1, pp. 1583--1601, 2020.

\bibitem{4908302}
J.~Nie, X.~Chen, and W.~Wang, ``Utility-{B}ased {R}esource {D}ynamic
  {A}llocation for {M}ixed {T}raffic in {W}ireless {N}etworks,'' in \emph{2009
  International Conference on Networks Security, Wireless Communications and
  Trusted Computing}, vol.~1, 2009, pp. 443--446.

\bibitem{5230848}
W.~Saad, Z.~Han, M.~Debbah, A.~Hjorungnes, and T.~Basar, ``Coalitional {G}ame
  {T}heory for {C}ommunication {N}etworks,'' \emph{IEEE Signal Processing
  Magazine}, vol.~26, no.~5, pp. 77--97, 2009.

\bibitem{8648014}
J.~Hu, Z.~Zheng, B.~Di, and L.~Song, ``Tri-{L}evel {S}tackelberg {G}ame for
  {R}esource {A}llocation in {R}adio {A}ccess {N}etwork {S}licing,'' in
  \emph{2018 IEEE Global Communications Conference (GLOBECOM)}, 2018, pp. 1--6.

\bibitem{9833469}
G.~O. Boateng, G.~Sun, D.~A. Mensah, D.~M. Doe, R.~Ou, and G.~Liu, ``Consortium
  {B}lockchain-{B}ased {S}pectrum {T}rading for {N}etwork {S}licing in 5{G}
  {R}{A}{N}: {A} {M}ulti-{A}gent {D}eep {R}einforcement {L}earning
  {A}pproach,'' \emph{IEEE Transactions on Mobile Computing}, pp. 1--15, 2022.

\bibitem{9594082}
G.~O. Boateng, D.~Ayepah-Mensah, D.~M. Doe, A.~Mohammed, G.~Sun, and G.~Liu,
  ``Blockchain-{E}nabled {R}esource {T}rading and {D}eep {R}einforcement
  {L}earning-{B}ased {A}utonomous {R}{A}{N} {S}licing in 5{G},'' \emph{IEEE
  Transactions on Network and Service Management}, vol.~19, no.~1, pp.
  216--227, 2022.

\bibitem{Sutton1998}
\BIBentryALTinterwordspacing
R.~S. Sutton and A.~G. Barto, \emph{Reinforcement {L}earning: {A}n
  {I}ntroduction}, 2nd~ed.\hskip 1em plus 0.5em minus 0.4em\relax The MIT
  Press, 2018. [Online]. Available:
  \url{http://incompleteideas.net/book/the-book-2nd.html}
\BIBentrySTDinterwordspacing

\bibitem{9322481}
Z.~Wang, Y.~Wei, F.~R. Yu, and Z.~Han, ``Utility {O}ptimization for {R}esource
  {A}llocation in {E}dge {N}etwork {S}licing {U}sing {D}{R}{L},'' in
  \emph{GLOBECOM 2020 - 2020 IEEE Global Communications Conference}, 2020, pp.
  1--6.

\bibitem{9444840}
D.~Shi, L.~Li, T.~Ohtsuki, M.~Pan, Z.~Han, and V.~Poor, ``Make {S}mart
  {D}ecisions {F}aster: {D}eciding {D}2{D} {R}esource {A}llocation via
  {S}tackelberg {G}ame {G}uided {M}ulti-{A}gent {D}eep {R}einforcement
  {L}earning,'' \emph{IEEE Transactions on Mobile Computing}, pp. 1--1, 2021.

\end{thebibliography}

\end{document}